\documentclass[aip,jap,reprint,graphicx]{revtex4-1}

\usepackage{graphicx, subfigure}% Include figure files
\begin{document}

\title[Numerical optimization of writer geometries for bit patterned magnetic recording]{Numerical optimization of writer geometries for bit patterned magnetic recording}% Force line breaks with \\
\thanks{Copyright 2014 American Institute of Physics. This article may be downloaded for personal use only. Any other use requires prior permission of the author and the American Institute of Physics. The following article appeared in "A. Kovacs et al., Journal of Applied Physics, 115, 17B704 (2014)" and may be found at \url{http://dx.doi.org/10.1063/1.4859055}.}

\author{A. Kovacs}
\email{alexander.kovacs@fhstp.ac.at}
\author{H. Oezelt}
\author{S. Bance}
\author{J. Fischbacher}
\author{M. Gusenbauer}
\author{F. Reichel}
\author{L. Exl}
\author{T. Schrefl}
\affiliation{
St. Poelten University of Applied Sciences, Matthias Corvinus Str. 15, 3100 St. Poelten, Austria%\\This line break forced with \textbackslash\textbackslash
}
\author{M. E. Schabes}
\affiliation{
HGST, 3403 Yerba Buena Road, San Jose, CA 9513
}%
\date{\today}% It is always \today, today,

\begin{abstract}
A fully-automated pole-tip shape optimization tool, involving write head geometry construction, meshing, micromagnetic simulation and evaluation, is presented. Optimizations have been performed for three different writing schemes (centered, staggered and shingled) for an underlying bit patterned media with an areal density of $2.12$ Tdots/in$^2$. Optimizations were performed for a single-phase media with 10 nm thickness and a mag spacing of 8 nm. From the computed write field and its gradient and the minimum energy barrier during writing for islands on the adjacent track, the overall write error rate is computed. The overall write errors are 0.7, 0.08, and \(2.8\times10^{-5}\) for centered writing, staggered writing, and shingled writing.
%
%\\
%Valid PACS numbers may be entered using the \verb+\pacs{#1}+ command.
\end{abstract}

\maketitle

\section{\label{sec:intro}Introduction}
Bit patterned magnetic recording poses many novel challenges, in terms of media manufacturing but also in terms of the recording physics \cite{Schabes2008}. Bit patterned media recording requires a localized write field. Good down-track
field gradients are important for island addressability. Good cross-track
field gradients are needed to support the high track density. The distribution of the write field strongly depends on the pole tip shape and shield distances. For best writer performance both the effective write field and the write field gradient should be maximized. In addition the effective write field should be small enough to avoid adjacent track erasure. Therefore, finding the best writer design is a constrained multi-objective optimization problem in a high dimensional configuration space, which grows in dimensions the more design parameters are taken into account. Multiple sweeps of a single design parameter are numerically too expensive to find the optimal solution. 
%Traditional design-of-experiment techniques may need significant a-priori
%knowledge of the search-space in order to locate the
%optimum design. 

We combine micromagnetic finite element simulations \cite{Schrefl2005} with a numerical optimization tool for multi-objective optimization\cite{Sherpa}. The combination of finite element analysis with optimization has a long tradition in the automotive industry \cite{Wang2004} and electrical engineering \cite{Parasiliti2012}. Fukuda and co-workers\cite{fukuda2012} simultaneously optimized writer and media parameters for granular perpendicular recording, using   a genetic algorithm together with a finite element static Maxwell solver and a micromagnetic solver.

  We focus on writer optimization for bit patterned media. Numerical optimization methods are iterative and require many evaluations of the objective function. In order to reduce the number of finite element micromagnetic field evaluations, we apply the response surface method \cite{Sherpa} that locally approximates the objective functions.

Bit patterned media is a candidate for extending magnetic data storage towards $10 $ Tb/in$^2$ and many papers have been published showing its potential \cite{Schabes2008, Richter2006a, Dong2012, Muraoka2011, Wood2000}. Writing on continuous granular media, where a bit cell is formed by a large group of grains, no loss of information appears if a few grains are not switched by the write field. But looking at bit patterned media, where each bit cell is formed by just one single island, we now have to assess if switching has occurred and introduce bit error rates\cite{Muraoka2011, Muraoka2008}. In order to analyze the performance, we can  use multiple micromagnetic simulations of bit patterned media ensembles \cite{Schabes2008} and count the write errors or a statistical approaches to compute the write error rate \cite{Muraoka2011}. In this paper we aim for the statistical approach. In section \ref{sec:method} of this paper we describe the recording head geometry, the media design and the iterative optimization process. Section \ref{sec:results} shows the optimized head structures for each writing scheme (centered, staggered and shingled writing), their performance and bit error rates.

\section{\label{sec:method}Method}
The optimization cycle 
%(shown in Figure \ref{fig:cycle})
consists of two major parts. The \textit{model calculation} constructs and analyzes a write head model according to a given set of design parameters, and the iterative \textit{optimization process}, which suggests new sets of design parameters based on the previous results.

The calculation of a model is performed by a Python script, which reads in given design parameters and produces a write head geometry accordingly. The model consists of a full write head structure with coils and a soft underlayer (Section \ref{sec:headdesign}). The script meshes the model, performs a micromagnetic simulation and extracts and evaluates the simulation results.
The algorithm 
%shown in Figure \ref{fig:cycle} 
can run fully automatically. Computer aided design is done with the software package Salome \cite{Salome}. Meshing with fine mesh near the pole tip (2.5 nm) is done with the mesh generation program Netgen\cite{Netgen}. A hybrid finite element /\ boundary element solver \cite{Schrefl2005}\ was used for write field calculation. After bringing the write head into a remanent state we apply an 80 mA coil current pulse with a rise time of 0.1 ns. After 2 ns we compute the write field below the saturated write head with a resolution of 2.5 nm. Evaluation of fields and gradients are done in the center of the target track and the center of the adjacent track.  

\subsection{\label{sec:headdesign}Head design}

\begin{figure}
        \begin{center}
        \subfigure[]{%
        \label{fig:xtparams}
        \includegraphics[scale=.25]{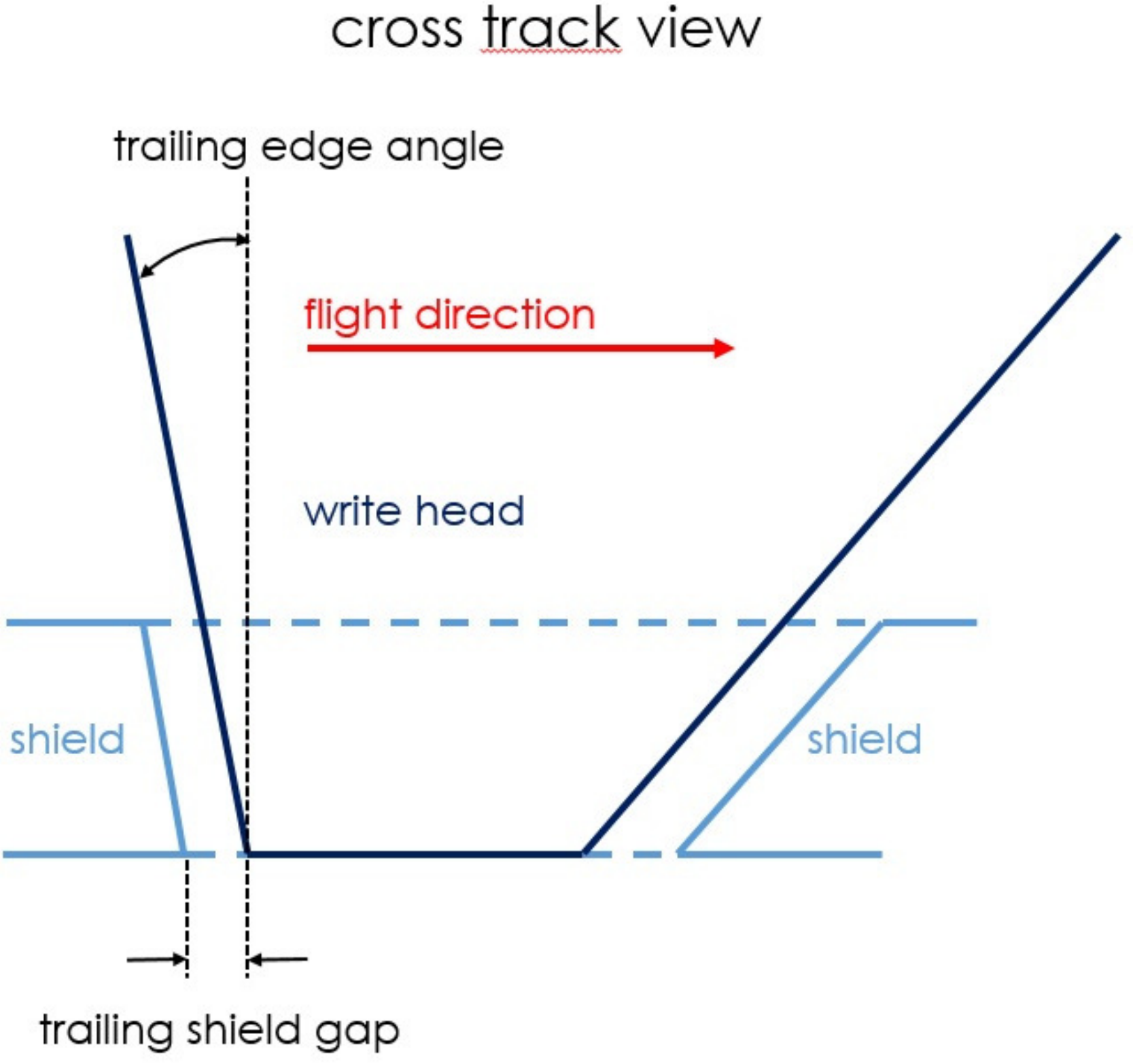}%
        }
        \subfigure[]{%
        \label{fig:dtparams}
        \includegraphics[scale=.25]{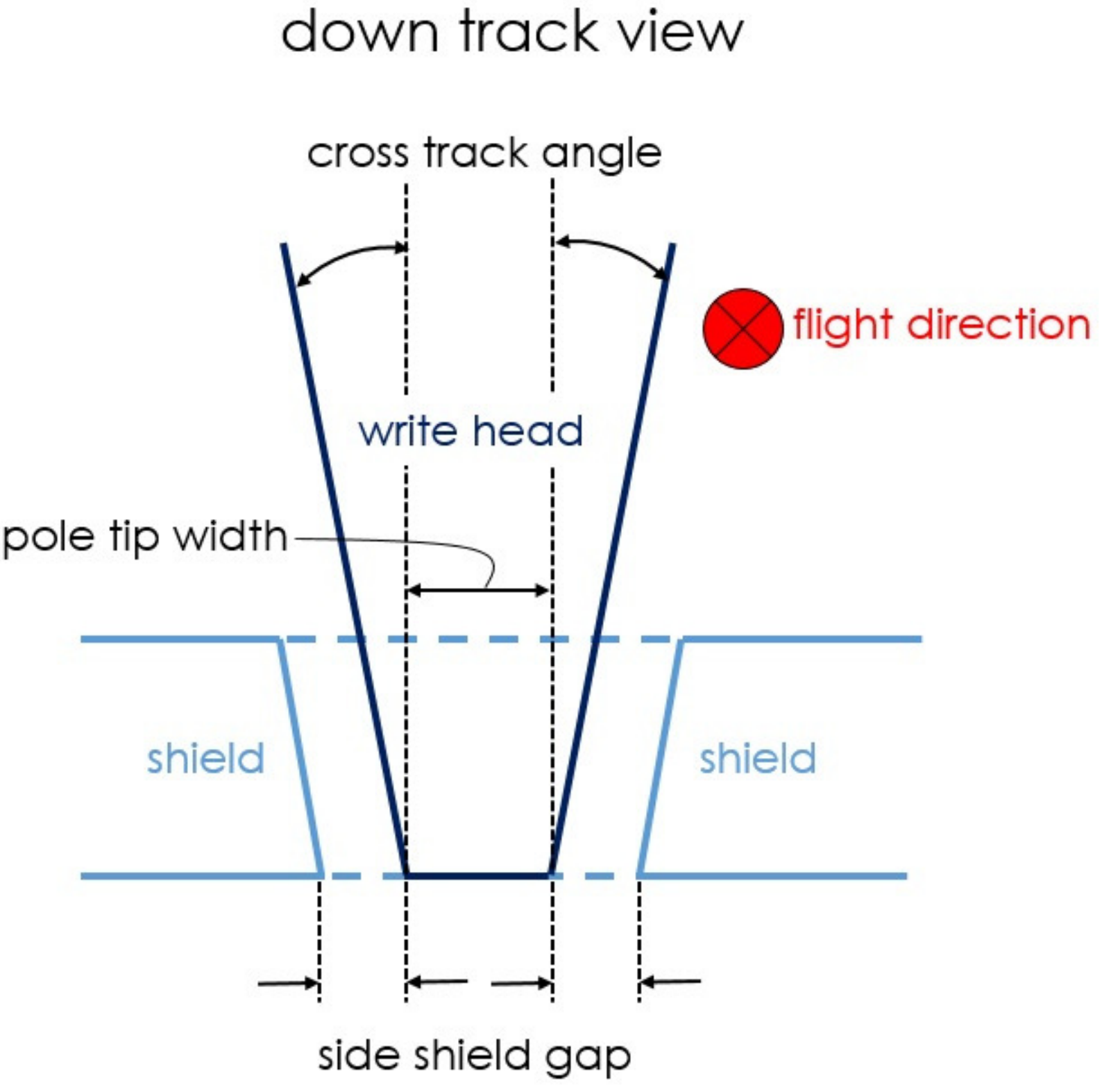}%
        }
        \caption{\label{fig:params}Cross track view \subref{fig:xtparams} and down track view \subref{fig:dtparams} of the pole tip and its design parameters. Side shield and trailing shield gaps are varied, as well as the cross track and trailing edge angles.}%
        \end{center}
\end{figure}

The write head geometry is constructed according to design parameter ranges (see figure \ref{fig:params}): trailing\ shield gap [5 nm, 20 nm], side shield gap [5 nm, 20 nm],  pole tip trailing edge angle [0,\(\pi/4\)], pole tip cross track angle [0,\(\pi/4\)], pole tip width [10 nm, 30 nm], and cross track offset [-10 nm, 10 nm]. The interval gives the possible range of a parameter. For shingled writing the pole tip width\ is fixed to 80 nm and a skewing angle of \(\pi/12\) is used. The write head is constructed with a helical coil with 4 turns. For the main pole we use a magnetic polarization of 2.4 T and an exchange constant of 20.15 pJ/m. The shields have a magnetic polarization of 2 T and an exchange of 13 pJ/m. The distance between the air bearing surface and soft under layer is 20 nm.

\subsection{\label{sec:mediadesign}Media design}
\begin{figure}
        \begin{center}
        \subfigure[Top view]{%
        \label{fig:media_tv}
        \includegraphics[scale=.16]{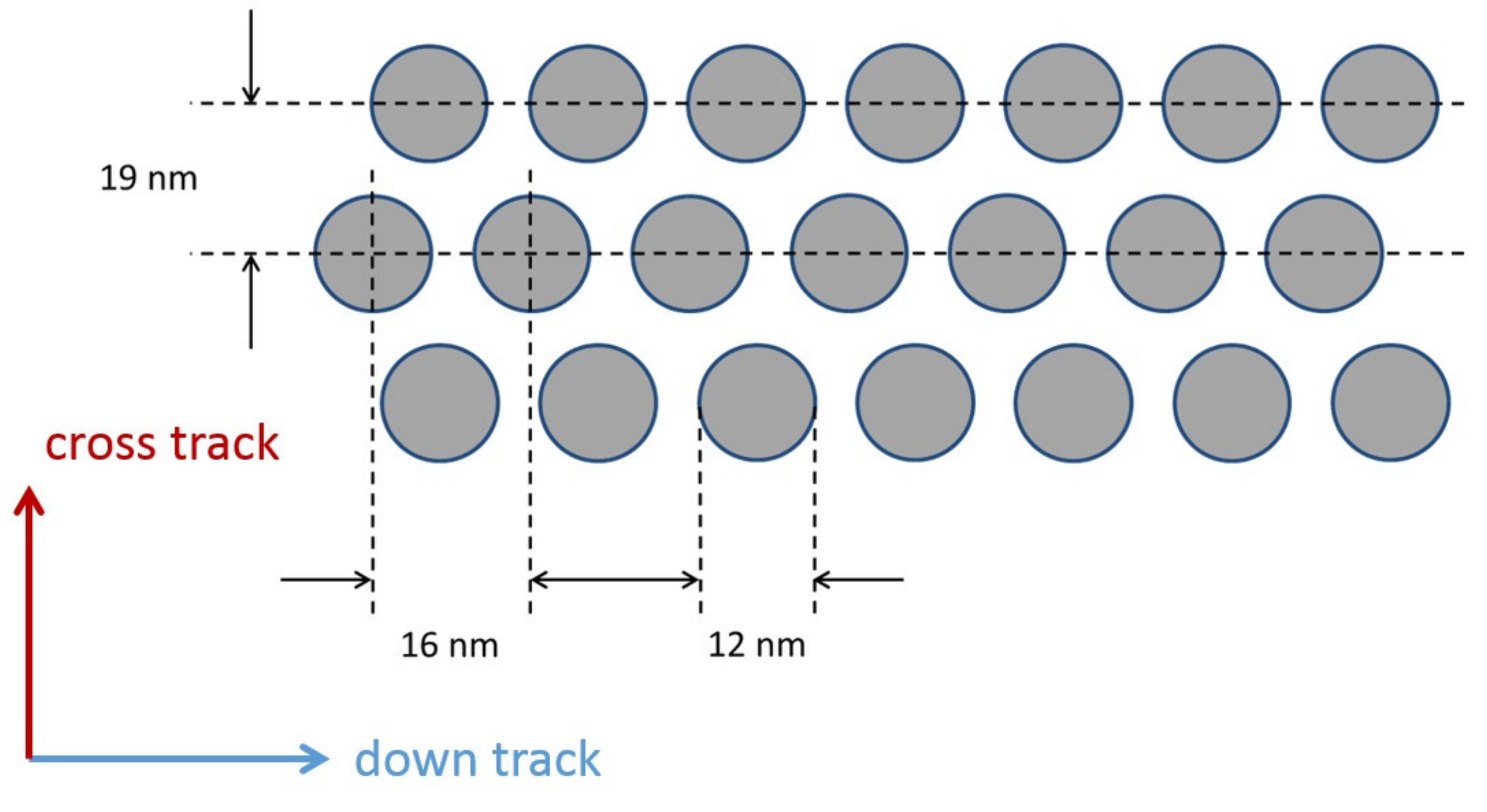}%
        }
        \subfigure[Side view]{%
        \label{fig:media_sv}
       \includegraphics[scale=.16]{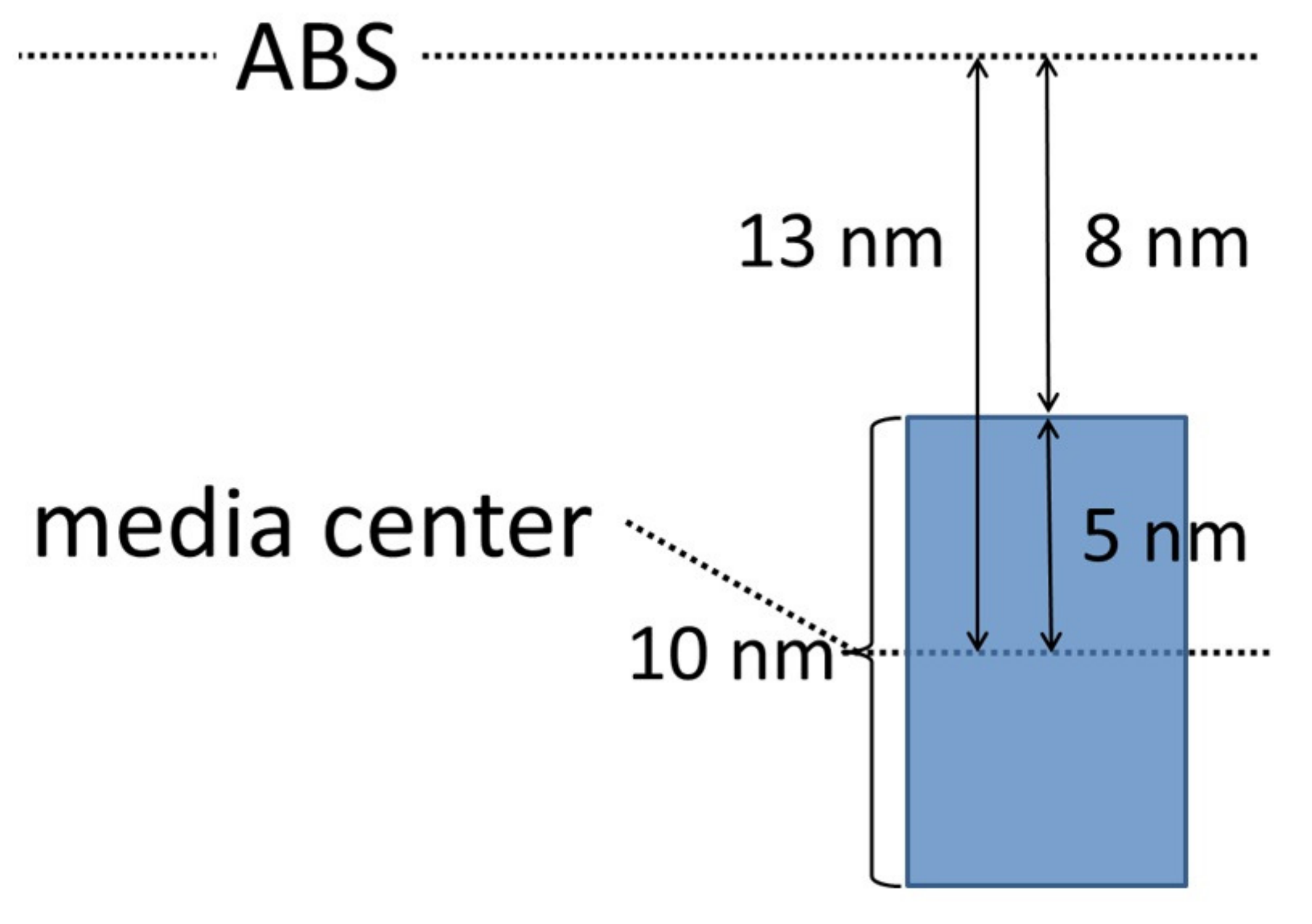}%
        }
        \caption{\label{fig:medialayout}Pseudo-hexagonal layout of a bit patterned single phase media. Cross track pitch is $19 \mathrm{nm}$, down track pitch is $16 \mathrm{nm}$ and dot diameter is $12 \mathrm{nm}$. Dot height is $10 \mathrm{nm}$, with a magnetic spacing of $8 \mathrm{nm}$. The magnetic polarization of the islands is $0.72$ T.  The anisotropy is adjusted to minimize  the total write error rate.    }%
        \end{center}
\end{figure}

While the pole tip shape and the shield distances vary, a predefined bit patterned media layout is used as shown in \ref{fig:medialayout}. Single phase cylindrical dots with a diameter of $12 \mathrm{nm}$ were used as the target media. Separated by a cross track pitch of $19 \mathrm{nm}$ and a down track pitch of $16 \mathrm{nm, which}$ gives us a media layout with an overall areal density at $2.12$ Tdots/in$^2$.

Design parameters were optimized for three different writing schemes: centered writing, staggered writing and shingled writing. 
%(Illustration of each writing scheme in figure \ref{fig:writingschemes}).
Centered writing is the classical writing scheme and focuses on one track center only. The staggered writing scheme \cite{Richter2007} gives the opportunity to increase the pole tip size so that the write field is focused above two tracks. The write head has to switch twice as fast as for the centered or shingled writing scheme.  The write head for shingled writing \cite{Wood2000} is built so that only one of its corners writes on a track.

\subsection{\label{sec:optimization}Optimization}

%The optimization software suggests a set of design parameters and performs the script execution (\textit{model calculation}). The optimizer reads in the results which represent the performance of the write head and suggests a new set of design parameters based on an intelligent search algorithm called MO-SHERPA \cite{Sherpa}. The objective functions for the optimization procedure are defined, so that (1) the effective field gradient, ${dH_{\mathrm{eff}}}{/dx},$ in down track direction and (2) the effective field  at the point of maximum field gradient, $H_{\mathrm{eff@maxgrad}},$ are maximized.  Furthermore, we keep the effective field along the adjacent track below a certain value with a constraint in order to avoid adjacent track erasure. The iterative procedure is repeated until there is no significant improvement in the solution or a predefined number of evaluations (500 iterations) have been executed.
The optimization software suggests a set of design parameters and performs the script execution (\textit{model calculation}). The optimizer reads in the results which represent the performance of the write head and suggests a new set of design parameters based on an multi-objective search algorithm  \cite{Sherpa}, which uses a non-dominated sorting scheme to rank designs. The two objective functions are $f_1={dH_{\mathrm{eff}}}{/dx}$ (maximize field gradient) and $f_2=H_{\mathrm{eff}}\left(x_{\mathrm{max}}\right)$, $x_{\mathrm{max}}$ is the position where ${dH_{\mathrm{eff}}}{/dx}$ reaches its maximum. Furthermore, we keep the effective field along the adjacent track below a certain value with a constraint in order to avoid adjacent track erasure. The iterative procedure is repeated until there is no significant improvement in the solution or a predefined number of evaluations (500 iterations) have been executed.

\subsection{\label{sec:evaluation}Evaluation of write error rate}
\begin{figure}
        \begin{center}
        \label{fig:conclusio_right}
        \includegraphics[width=5cm]{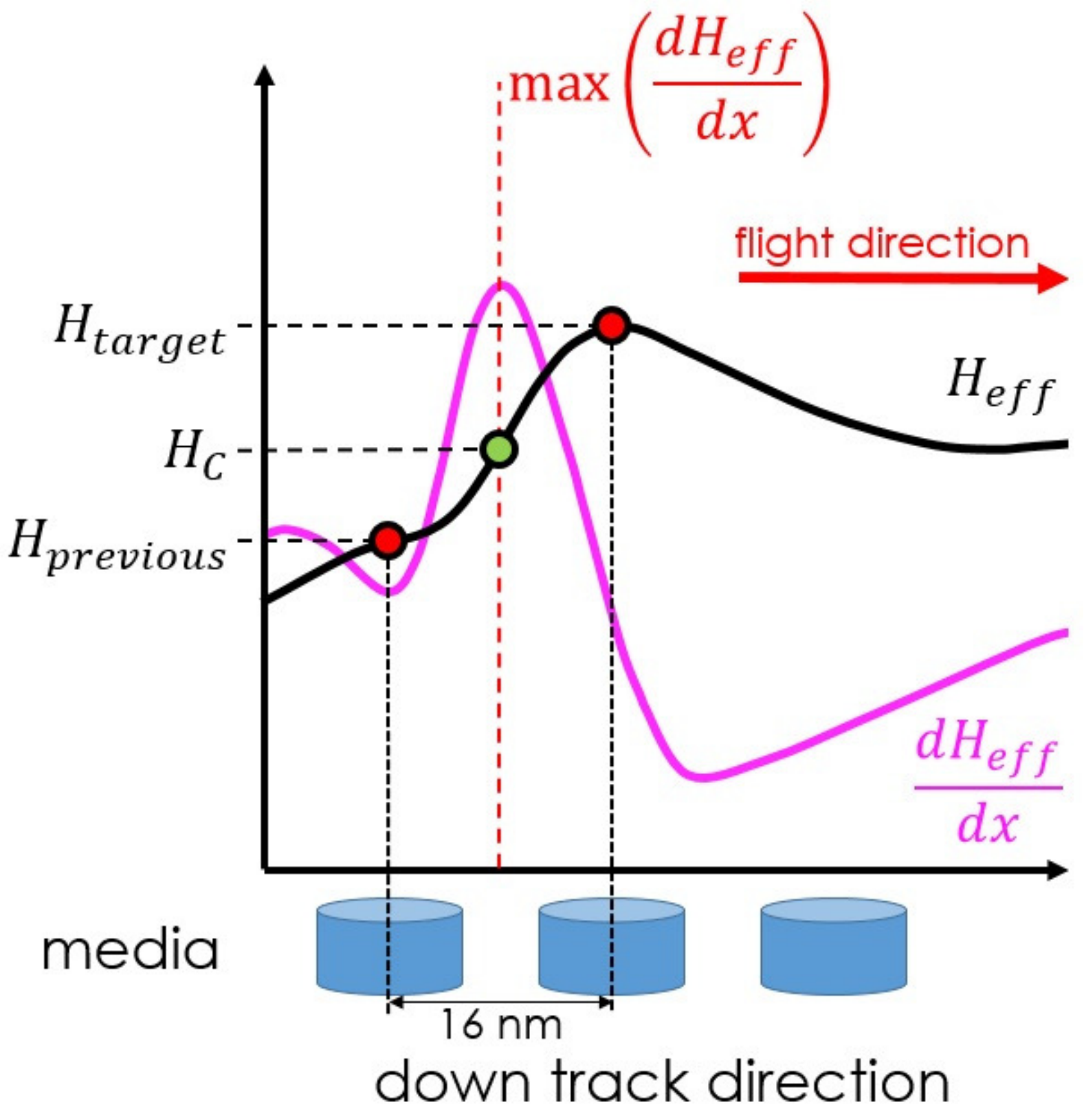}
\caption{\label{fig:conclusion}Field profile of an optimized write head. }%
        \end{center}
\end{figure}

In order to calculate the total write error rate \cite{Dong2012, Muraoka2011} we  add  the following contributions: (1) not switching the target bit, (2) switching an already written bit, (3) switching a bit on the adjacent track.
We approximate the  coercive field of an island with its anisotropy field \(H_\mathrm{C} \approx H_\mathrm{A}\) (the shape anisotropy is small).   At this point  the write field profile preferably has the highest field gradient.  From the working point we go along the track exactly half of the down track pitch in negative and positive down track direction and gather the  fields $H_{\mathrm{prev}}$ and $H_{\mathrm{targ}}$ (see Fig. \ref{fig:conclusion}) . The bit error rate  (1) is $BER_{\mathrm{targ}}=\frac{1}{2}\left(1-erf\left(\frac{H_{\mathrm{targ}}-H_C}{\sqrt{2}\sigma_{\mathrm{SW}}}\right)\right),$ where $\sigma_{\mathrm{SW}}=\sqrt{\sigma_{H,pos}^2+\sigma_H^2+\sigma_K^2}$ and $\sigma_{H,\mathrm{pos}}=\sigma_{\mathrm{pos}}\frac{dH_{\mathrm{eff}}}{dx}$. Shield effects\cite{Wood2000} reduce the interaction field distribution, \(\sigma_{H}\), from 0.032 T to 0.02 T. The sigmas of the dot position and the anisotropy field are $\sigma_{\mathrm{pos}}=0.8$ nm and  $\sigma_K=0.05$ T. 
The error rate (2) is  $BER_{\mathrm{prev}}=\frac{1}{2}\left(1-erf\left(\frac{H_C-H_{\mathrm{prev}}}{\sqrt{2}\sigma_{\mathrm{SW}}}\right)\right)$.
For the third case (3) we compute the thermally induced adjacent track erasure \cite{Dean2008}. We place an island into the precomputed write field on the adjacent track at the position with the highest effective field and compute the life time of the bit, $\tau=\frac{1}{f_0}\exp\left(E_B\right),$ where $E_B$ is the energy barrier computed with the nudged elastic band method and an attempt frequency\cite{Dean2008} of $1.3\times10^{11}$ Hz. We assume a field exposure time during writing of \(t_{\mathrm{write}}=1\)~ns. The number of passes before erasure is  $\tau/{t_{\mathrm{write}} }$ and  $BER_{\mathrm{adj}}=t_{\mathrm{write}}/\tau.$ The total write error rate is $BER_{\mathrm{tot}}=BER_{\mathrm{targ}}+BER_{\mathrm{prev}}+BER_{\mathrm{adj}}$.

\section{\label{sec:results}Results and Discussion}
 In table \ref{tab:ber} we show bit error rates for each writing scheme where we only changed the coercive field of the media by shifting the working point in the field profile. Thermally induced adjacent track erasure is dominating for centered and staggered writing. Higher anisotropy media (higher \(H_{C}\)) improves $BER_{\mathrm{adj}}$ but reduces the writeability at the target bit.   Table \ref{tab:bestdesignparams} shows the optimal design parameters of each head. 

 The results show that bit patterned media recording on single phase islands and a magnetic spacing of 8 nm can only be achieved with shingled writing. For shingled writing an optimal write field profile (see Fig. 3) was found. Through optimization the point of maximum field gradient moved towards the point of maximum write field. Thus  both on-track errors and cross track errors were reduced. All errors have the same order of magnitude.

\begin{table}
\caption{\label{tab:ber} }
\begin{tabular}{c | c | c | c | c | c}
writing  & 
$H_\mathrm{A}$ [T] & 
$\frac{dH_\mathrm{eff}}{dx}$  & & 
& 
\\
scheme &
& $\left[\frac{\mathrm{mT}}{\mathrm{nm}}\right]$ & 
$BER_{\mathrm{targ}}$ & $BER_{\mathrm{prev}}$ & $BER_{\mathrm{adj}}$\\ \hline
centered & 0.75 & 33 & 5.1$\times10^{-5}$ & 1.3$\times10^{-5}$ & $>1$ \\ \hline
centered & 0.90 & 27 & 1.9$\times10^{-2}$ & 9.4$\times10^{-7}$ & 5.5$\times10^{-1}$ \\ \hline
%centered & XX & XX & XX & XX & XX \\ \hline
staggered & 0.82 & 38 & 1.3$\times10^{-6}$ & 4.7$\times10^{-6}$ & $>1$ \\ \hline
staggered & 1.07 & 27 & 2.9$\times10^{-2}$ & 7.2$\times10^{-8}$ & 5.2$\times10^{-2}$ \\ \hline
%staggered & 0.97 & 34 & 4.9$\times10^{-5}$ & 3.0$\times10^{-9}$ & XX \\ \hline
shingled & 1.00 & 35 & 1.2$\times10^{-5}$ & 1.1$\times10^{-5}$ & 5.5$\times10^{-6}$\\ \hline
shingled & 1.10 & 34 & 1.4$\times10^{-4}$ & 2.3$\times10^{-6}$ & 2.7$\times10^{-8}$\\ %\hline
%shingled & 1.15 & 33 & 1.1$\times10^{-4}$ & 8.9$\times10^{-8}$ & 2.96$\times10^{-7}$
\end{tabular}
\end{table}

\begin{table}
\caption{\label{tab:bestdesignparams} }
\begin{tabular}{c | c | c | c | c | c }
writing & 
trailing & 
side & 
trailing & 
cross  & 
pole  
\\ 
scheme & 
shield & 
shield & 
edge & 
track & 
tip 
\\ 
& 
gap [nm] & 
gap [nm] & 
angle [$^\circ$] & 
angle [$^\circ$] & 
width [nm] 
\\ \hline
centered & 9 & 16 & 20 & 15 & 14 \\ \hline
staggered & 13 & 11 & 35 & 10 & 29 \\ \hline
shingled & 14 & 12 & 45 & 5 & 80 
\end{tabular}
\end{table}

\begin{acknowledgments}
Work supported by ASTC IDEMA and the Austrian Science Fund (I821-N16).
\end{acknowledgments}

%\bibliography{ct08}


\end{document}